\documentclass[english,aip,reprint]{revtex4-1}
\usepackage{float}
\usepackage{amsmath}
\usepackage{wrapfig}
\usepackage{graphicx}
\usepackage{epstopdf}
\usepackage{tikz}
\usepackage{epsfig}
\makeatother
\begin{document}
\title{Generalized model for the diffusion of solvents in glassy polymers: From Fickian to super Case II}

\author{Jiayuan Miao}
\affiliation{
Department of Physics, Case Western Reserve University,
  Cleveland, Ohio 44106, USA}
\author{Mesfin Tsige}
\affiliation{
Department of Polymer Science, The University of Akron,
  Akron, Ohio 44325, USA}
\author{Philip L. Taylor}
\email[Author to whom correspondence should be addressed. Electronic mail: ]{taylor@case.edu}
\affiliation{
Department of Physics, Case Western Reserve University,
  Cleveland, Ohio 44106, USA}

\begin{abstract}

The diffusion of small solvent molecules in glassy polymers may take on a variety of different forms.  Fickian, anomalous, Case II, and super Case II diffusion have all been observed, and theoretical models exist that describe each specific type of behavior.  Here we present a single generalized kinetic model capable of yielding all these different types of diffusion on the basis of just two parameters.  The principal determinant of the type of diffusion is observed to be a dimensionless parameter, $\gamma$, that describes the influence of solvent-induced swelling in lowering the potential barriers separating available solvent sites.   A second  parameter, $\eta$, which characterizes the effect of the solvent in reducing the potential energy of a solvent molecule when at rest at an available site, only influences the type of diffusion to a lesser extent.  The theoretical analysis does not include any effects that are explicitly non-local in time, an example of which is the inclusion of polymer viscosity in the Thomas-Windle model; it thus represents a variant of Fick's second law utilizing a concentration-dependent diffusivity when $\eta$ is small.   To check the significance of time-delayed swelling, a simulation was performed of a modified model that contained a history-dependent term.   The results were found to be very similar to those in the time-local model.

\end{abstract}
\maketitle

\section{Introduction}
\label{section-introduction} 

In the study of the diffusion of small molecules into polymeric materials, several different types of behavior have been observed. They are classified according to the equation: $M_t=kt^n$, where $M_t$ is the amount of solvent absorbed per unit area of polymer at an elapsed time $t$ after the two have been placed in contact, and $k$ is a constant. When the exponent $n=1/2$, it is the familiar Fickian diffusion; when $n=1$, it is known as Case II diffusion, and when $1/2<n<1$ it is termed anomalous diffusion. \cite{masaro}
Jacques {\em et al.} \cite{jac} found that there is an acceleration of diffusion when the two fronts originating from the two opposite sides of a sample slab meet during Case II diffusion, and called it super Case II diffusion.

Predicting the detailed nature of the diffusion of the penetrant has turned out to be a challenging problem, and most of the proposed theoretical models apply only to a specific subinterval of the diffusion realm. For example, Thomas and Windle proposed a widely discussed model for Case II diffusion. \cite{thom} Qian and Taylor made an effort to explain the crossover from Case II to Fickian behavior using two separate equations. \cite{qian} Hansen has also tried to incorporate different kinds of diffusion behavior by modifying the boundary conditions of the diffusion equation and using diffusion coefficients that are exponentially dependent on the concentration. \cite{hansen} Wilmers and Bargmann also used an exponentially concentration-dependent diffusion coefficient while incorporating a dual-phase-lag concept to Fick's first law and derived a promising description of Case II behavior. \cite{wilmers}

In this paper, we present a single model that exhibits all the observed types of diffusion. We begin by noting that solvent-induced swelling can affect the diffusivity in two distinct ways. One consequence of the swelling is a decrease in the potential energy of solvent molecules in the interstices of the polymer matrix. A second consequence is that the height of the potential-energy barriers that must be surmounted is reduced. We explore the respective effects of these phenomena in detail, using first a theoretical analysis and then a numerical simulation. We find the latter effect to be the principal determinant of the diffusive behavior. With an appropriate parameter set, all the different types of diffusion mentioned above can be reproduced.

\section{Theoretical Analysis}

\subsection{The background}

In the simplest possible model of diffusion, the diffusing particles execute a random walk within the interstices of a fixed network of obstructions. On a macroscopic scale, in which a local concentration $\varphi ({\bf r})$ of diffusant can be defined, one finds that $\varphi$ obeys Fick's law, \cite{crank}
\begin{equation} \label{eq1}
\frac{\partial \varphi}{\partial t}=D\nabla^2\varphi,
\end{equation}
and the process is known as Fickian diffusion. The development with time of the function $\varphi ({\bf r},t)$ is entirely determined by the initial conditions and the constant diffusivity $D$.

The process by which an organic solvent diffuses into a real polymer is much more complex than the simple Fickian picture. The most important difference is that the presence of the solvent modifies the diffusion process. This can occur through a variety of mechanisms, the simplest of which is that entry of solvent may cause the polymer to swell, thus widening the channels through which the solvent molecules must pass. These processes have the effect of making the diffusivity $D$ a function of solvent concentration. The non-linear diffusion equation that results is
\begin{equation} \label{eq2}
\frac{\partial \varphi}{\partial t}=\nabla \cdot \left[D(\varphi)\nabla \varphi\right]
\end{equation}
when other complications are ignored. 

\begin{figure}[!htb]
\begin{center}
\includegraphics[scale=0.3]{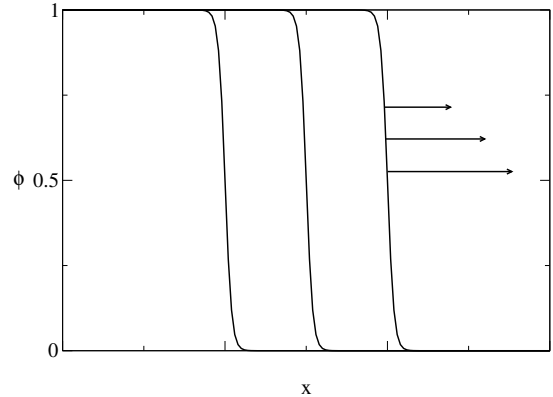}
\caption{A schematic of the penetrant concentration ($\phi=\varphi/\varphi_0$) as predicted by Eq.~(\ref{eq2}) with $D=D_0/(\varphi_0-\varphi)$. The arrows show the direction in which the front is advancing.}
\label{solitary}
\end{center}
\end{figure}

Although Eq.~(\ref{eq2}) is an oversimplification of a real polymer diffusion problem, it can provide good insight into the type of non-linear diffusion that is possible. An example would be the exponential dependence of $D$ on $\varphi$ first proposed by Prager, \cite{prager} which has been the subject of a recent study by Marais {\em et al.} \cite{marais} In another special case, $D=D_0/(\varphi_0-\varphi)$ with $D_0$ and $\varphi_0$ constants. We then have a diffusivity that is inversely proportional to the polymer density, and Eq.~(\ref{eq2}) has a solution for diffusion in one dimension of the form
\begin{equation} \label{eq3}
\varphi (x,t)=\varphi_0 \left(e^{q(x-vt)} + 1\right)^{-1},
\end{equation}
with $v$ a velocity and $q=\varphi_0 v/D_0$. This represents a solitary wave in which the solvent advances into the polymer as a front moving with constant speed $v$, as illustrated in Fig.~\ref{solitary}. It thus represents an example of what has been termed Case II diffusion, \cite{thomas1,thom,alfrey,rossi,thomas3,lasky1,lasky2,hui1,hui2,mills} in which a glassy polymer is transformed to a more porous, rubbery state by the intrusion of solvent. The model of idealized Case II diffusion, in which the speed of the advancing front is constant, can only be an approximation to the real situation, in which there will always be some slowing of the rate at which the front advances. The model does, however, serve an important function in indicating that a study of the form of the concentration-dependent diffusivity $D(\varphi)$ can be a first step in developing a realistic macroscopic theory with good predictive power. We note, however, that the speed $v$ of the diffusion front is not determined by Eq.~(\ref{eq2}) and Eq.~(\ref{eq3}), which contain no parameter having the dimensions of a velocity.

\subsection{The model}

We consider a model of a {\em random walker}, for which the direction of motion of a solvent molecule from one site to an adjacent site is completely independent of the  direction of the previous motion by which the molecule arrived at that site. This is used because, in molecular dynamics simulations, one sees a solvent molecule oscillating many times within a cavity before making a transition to a new site. The momentum with which it arrives at a site is thus lost, and the direction of the next jump to another site is mostly uncorrelated with the direction of the previous one. There may remain a small negative correlation due to the swelling of a previously occupied site. 

\begin{figure}[!htb]
\begin{center}
\includegraphics[scale=0.5]{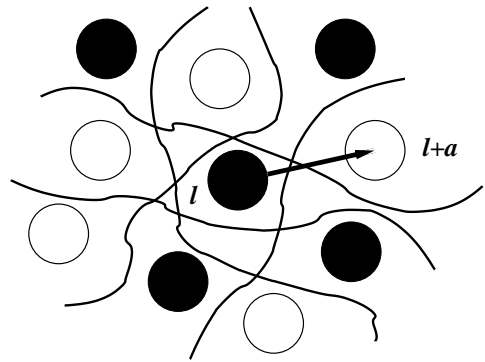}
\caption{Within the polymer network there are some sites occupied by solvent molecules (shown as filled circles) and some empty (open circles).}
\label{random}
\end{center}
\end{figure}

This model is illustrated in Fig.~\ref{random}. Within the polymer network there are sites at positions ${\boldsymbol l}$ at which solvent molecules may reside. A particular configuration  will have some sites occupied (shown as filled circles) and some empty. We write the probability of occupancy of site ${\boldsymbol l}$ as $\phi_l$. This will be related to the local concentration $\varphi$ of solvent through the equation $\varphi=\rho_0 \phi_l$, with $\rho_0$ the concentration of sites. We then examine the current {\boldmath $J$} of solvent flowing between sites ${\boldsymbol l}$ and ${\boldsymbol l}+{\boldsymbol a}$, with ${\boldsymbol a}$ the displacement between a cavity and one of its nearest neighbors. 

\begin{figure}[!htb]
\begin{center}
\includegraphics[scale=0.5]{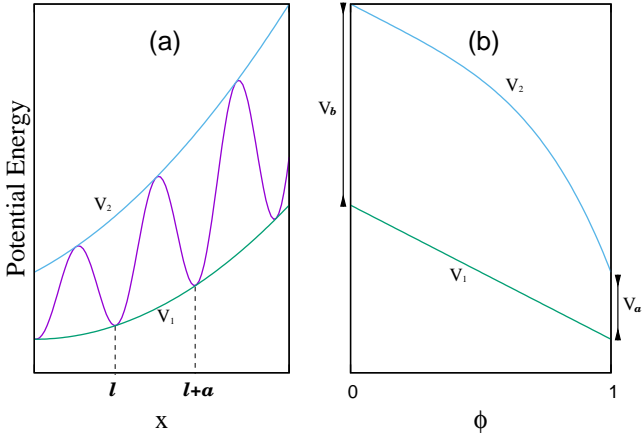}
\caption{The potential energy profile near the site ${\boldsymbol l}$ is shown schematically as a function of position (a) and of local concentration (b). }
\label{barrier}
\end{center}
\end{figure}

For a solvent molecule to make a transition from ${\boldsymbol l}$ to ${\boldsymbol l}+{\boldsymbol a}$ we must have site ${\boldsymbol l}$ occupied and site ${\boldsymbol l}+{\boldsymbol a}$ empty. There is thus a factor of $\phi_l (1-\phi_{l+a})$ in the expression for the current. There will also be an attempt frequency $\nu$ and an Arrhenius factor, $e^{-\beta V}$,  for the probability of success of the transition, with $\beta=1/k_{\rm B}T$, where $k_{\rm B}$ is Boltzmann's constant and $T$ is the absolute temperature. The quantity $V$ is an energy barrier height, which will differ for a transition in the forward direction from that for a reverse transition. The energy landscape is depicted in Fig.~\ref{barrier}(a). The potential energy at the local sites varies with solvent concentration, and hence with position, as $V_1$, while the potential at the summit of the barriers varies as $V_2$.
We can thus write the barrier $V$ for forward and reverse transitions as $V_2({\boldsymbol l}+\frac{{\boldsymbol a}}{2}) - V_1({\boldsymbol l})$ and $V_2({\boldsymbol l}+\frac{{\boldsymbol a}}{2}) - V_1({\boldsymbol l}+{\boldsymbol a})$, respectively. We define 
\begin{equation} \label{eqv0}
V_2\left({\boldsymbol l}+\frac{{\boldsymbol a}}{2}\right)-\frac{1}{2}\left[V_1({\boldsymbol l})+V_1({\boldsymbol l}+{\boldsymbol a})\right]\equiv V_0\left({\boldsymbol l}+\frac{{\boldsymbol a}}{2}\right)
\end{equation}
and
\begin{equation}
V_1({\boldsymbol l}+{\boldsymbol a})-V_1({\boldsymbol l})\equiv W\left({\boldsymbol l}+\frac{{\boldsymbol a}}{2}\right).
\end{equation}
When we sum the contributions from forward and reverse transitions we find the current ${\boldsymbol J}$ of solvent flowing between sites ${\boldsymbol l}$ and ${\boldsymbol l}+{\boldsymbol a}$ as
\begin{equation} \label{eq4}
\begin{split}
\mbox{\boldmath $J$}_{{\boldsymbol l},{\boldsymbol l}+{\boldsymbol a}}= & \mbox{\boldmath $a$} \nu e^{-\beta V_0} [\phi_l(1-\phi_{l+a})e^{-\frac{1}{2}\beta W}\\
& -\phi_{l+a}(1-\phi_{l})e^{\frac{1}{2}\beta W}],
\end{split}
\end{equation}
where $V_0$ and $W$ are evaluated at position ${\boldsymbol l}+\frac{{\boldsymbol a}}{2}$. The small difference $W$ in barrier height can be written as 
\begin{equation} \label{eq5}
W=-K(\phi)\mbox{\boldmath $a$}\cdot\nabla\phi, 
\end{equation}
where 
\begin{equation}
K(\phi)=-\frac{dV_1}{d\phi},
\end{equation}
since it can reasonably be assumed to depend linearly on the concentration gradient in all but the most extreme circumstances. In our present calculation, we assume that $K$ is a non-negative constant.
For concreteness, we also take a simple model for the form of $V_0(\phi)$. We expect the presence of solvent to have the most marked effect on lowering the potential barrier $V_0$ when $\phi$ is close to unity; a small concentration of solvent will only have a marginal effect on swelling. It is only when $\phi$ is largest that the polymer matrix will be loosened sufficiently to permit a uniform expansion. This indicates that $V_0(\phi)$ is a monotonically decreasing concave function on $[0,1]$, such as $V_0(\phi)=A-B\phi^r$ ($A$, $B$ and $r$ are positive numbers, and $r\ge2$). Taking an example among the variety of possible forms of $V_0$, the relation that we adopt to model this behavior is
\begin{equation} \label{v2eq6}
V_0(\phi)=V_b - (V_b-V_a)\phi^4,
\end{equation}
where $V_b$ is the potential barrier in the absence of solvent ($\phi=0$), while $V_a$ is the lowered barrier at saturation ($\phi=1$). The form of $V_1(\phi)$ and $V_2(\phi)$ are then as shown in Fig.~\ref{barrier}(b).

At this point we introduce two dimensionless parameters, $\gamma$ and $\eta$, to characterize the dependence of the potential-energy landscape on the solvent occupancy $\phi$. The first parameter is defined as 
\begin{equation}
\gamma \equiv \beta(V_b-V_a).
\end{equation}
As we see from Fig.~\ref{barrier}(a) and Eqs.~(\ref{eqv0}) and (\ref{v2eq6}), this measures the strength of the dependence of the barrier height on $\phi$. The second parameter is 
\begin{equation}
\eta \equiv \frac{1}{2}\beta K =-\frac{1}{2}\beta\frac{dV_1}{d\phi}.
\end{equation}
This is a measure of the dependence of the potential energy at a local minimum on $\phi$, as is also seen from Fig.~\ref{barrier}(a).

We can now rewrite Eq.~(\ref{eq4}), which expresses the current $\mbox{\boldmath $J$}_{{\boldsymbol l},{\boldsymbol l}+{\boldsymbol a}}$ in terms of the function $V_0(\phi)$ and $W(\phi)$, into a form in which the current depends on the dimensionless  parameters $\gamma$ and $\eta$. We interpret the value of $\phi$ appearing in $V_0(\phi)$ and $W(\phi)$ as $(\phi_l+\phi_{l+a})/2\simeq\phi_{l+\frac{a}{2}}$ and find
\begin{equation} \label{v2eq7}
\begin{split}
\mbox{\boldmath $J$}_{{\boldsymbol l},{\boldsymbol l}+{\boldsymbol a}}=&\mbox{\boldmath $a$} \nu e^{-\beta V_b} e^{\gamma \phi_{l+\frac{a}{2}}^4}[\phi_l(1-\phi_{l+a})e^{\eta(\phi_{l+a}-\phi_l)} \\
& -\phi_{l+a}(1-\phi_{l})e^{-\eta(\phi_{l+a}-\phi_{l})}] \\
=&\mbox{\boldmath $a$} \nu e^{-\beta V_b} e^{\gamma \phi_{l+\frac{a}{2}}^4}\{(\phi_l-\phi_{l+a})\cosh[\eta(\phi_{l+a}-\phi_l)] \\
&+(\phi_l+\phi_{l+a}-2\phi_l \phi_{l+a})\sinh[\eta(\phi_{l+a}-\phi_l)]\}.
\end{split}
\end{equation}
Conservation of matter tells us that: 
\begin{equation} \label{v2eq11}
\frac{\partial \phi}{\partial t}=-\nabla \cdot \mbox{\boldmath $J$},
\end{equation}
of which the discretized form is (considering only diffusion in one dimension)
\begin{equation}  \label{v2eq12}
\phi_l(t+\Delta t)-\phi_l(t)=\frac{\Delta t}{a}\left(J_{l-a,l}-J_{l,l+a}\right).
\end{equation}

\subsection{Results}

The combination of Eq.~(\ref{v2eq12}) with Eq.~(\ref{v2eq7}) was solved numerically. The initial and boundary conditions are:
\begin{equation}
\phi(x,t)|_{t=0}=0, \space (x>0)
\end{equation}
\begin{equation} \label{eq15}
\phi(x,t)|_{x=0}=1,
\end{equation}
\begin{equation} \label{eq17}
\phi(x,t)|_{x\to\infty}=0.
\end{equation}
To minimize the calculation time while keeping the solution convergent, the prefactor $\frac{\Delta t}{a}$ was set to $2.5\times10^{-6}$. 

Results for various values of $\gamma$ and $\eta$ were obtained as time-varying concentration profiles and as plots of mass intake as a function of time. In one limit, when $\gamma=0$ and $\eta=0$, Eq.~(\ref{v2eq12}) becomes Eq.~(\ref{eq1}), which is Fick's second law, and the mass intake varies  as $t^{1/2}$. In another limit, when $\gamma\gg1$, the factor $e^{-\beta V_0}$ has a shape similar to $D=D_0/(\varphi_0-\varphi)$, and we expect Eq.~(\ref{v2eq12}) will give us a solitary wave solution similar to Eq.~(\ref{eq3}), for which the mass intake varies as $t$. This suggests that the mass-time curves calculated for different values of $\gamma$ and $\eta$ might be fitted to a function of the form $f(t)=kt^n$, with $k$ and $n$ the fitting parameters, to determine the nature of the diffusion mechanism as discussed below.

\begin{figure*}[!htb]
\begin{center}
\includegraphics[scale=0.85]{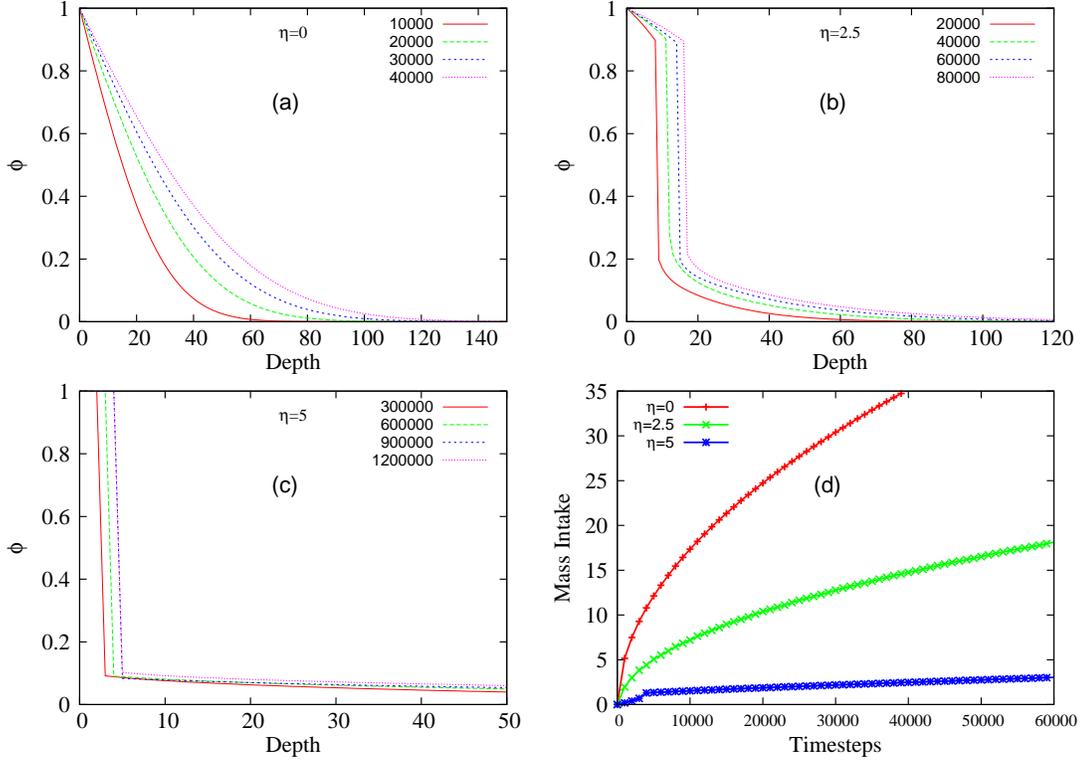}
\caption{Calculations performed with $\gamma$=0. (a)-(c): Concentration profiles at equally spaced time steps when $\eta=0, 2.5, 5.0$, respectively. (d): Mass intake as a function of time for different values of $\eta$: $0, 2.5, 5.0$.}
\label{numfickian0}
\end{center}
\end{figure*}

\begin{figure*}[!htb]
\begin{center}
\includegraphics[scale=0.85]{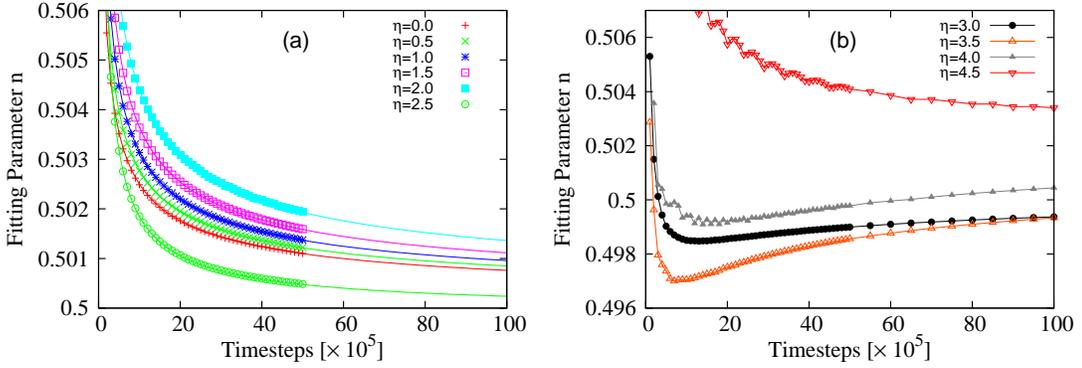}
\caption{The variation of fitting parameter $n$ with respect to diffusion time $t$ when $\gamma = 0$. (a) For $\eta = 0, 0.5, 1.0, 1.5, 2.0, 2.5$, the fitting functions are also shown as solid lines. (b) For $\eta = 3.0, 3.5, 4.0, 4.5$, the solid lines are a guide to the eye only.}
\label{neta}
\end{center}
\end{figure*}

The results for $\gamma=0$ are shown in Fig.~\ref{numfickian0} as concentration profiles at several times and for three different values of $\eta$ in Figs.~\ref{numfickian0}(a), \ref{numfickian0}(b), and \ref{numfickian0}(c), while the time variation of the mass intake is shown in Fig.~\ref{numfickian0}(d) for the same $\eta$ values.  The fitting parameter $n$ varies with the elapsed time $t$; it is shown in Fig.~\ref{neta}. For $\eta = 0, 0.5, 1.0, 1.5, 2.0$, and $2.5$ the $n$-$t$ curves are well fitted by a function of the form $n=n_\infty+bt^{-c}$, as can be seen in Fig.~\ref{neta}(a). This allows an accurate estimate of $n_\infty$, the long-time limit of $n$, as 0.49992, 0.49998, 0.50002, 0.50006, 0.50006 and 0.49989 respectively for these values of $\eta$. The $n$-$t$ relations shown in Fig.~\ref{neta}(b) for $\eta = 3.0, 3.5, 4.0, 4.5$ could not be fitted well, but we can estimate values of $n_\infty$ as 0.49937, 0.49933, 0.50044 and 0.50279 by extending the diffusion time. The $n_\infty$ for $\eta = 5.0$, which is outside the range shown in Fig.~\ref{neta}, is calculated to be 0.49918. Generally speaking, the mass-time curves follow asymptotically a $t^{1/2}$ relation, characteristic of Fickian diffusion. The influence of $\eta$ on $n$ is significant only at an early stage in the diffusion process, and is negligible at long diffusion times. 

\begin{figure*}[!htbp]
\begin{center}
\includegraphics[scale=0.85]{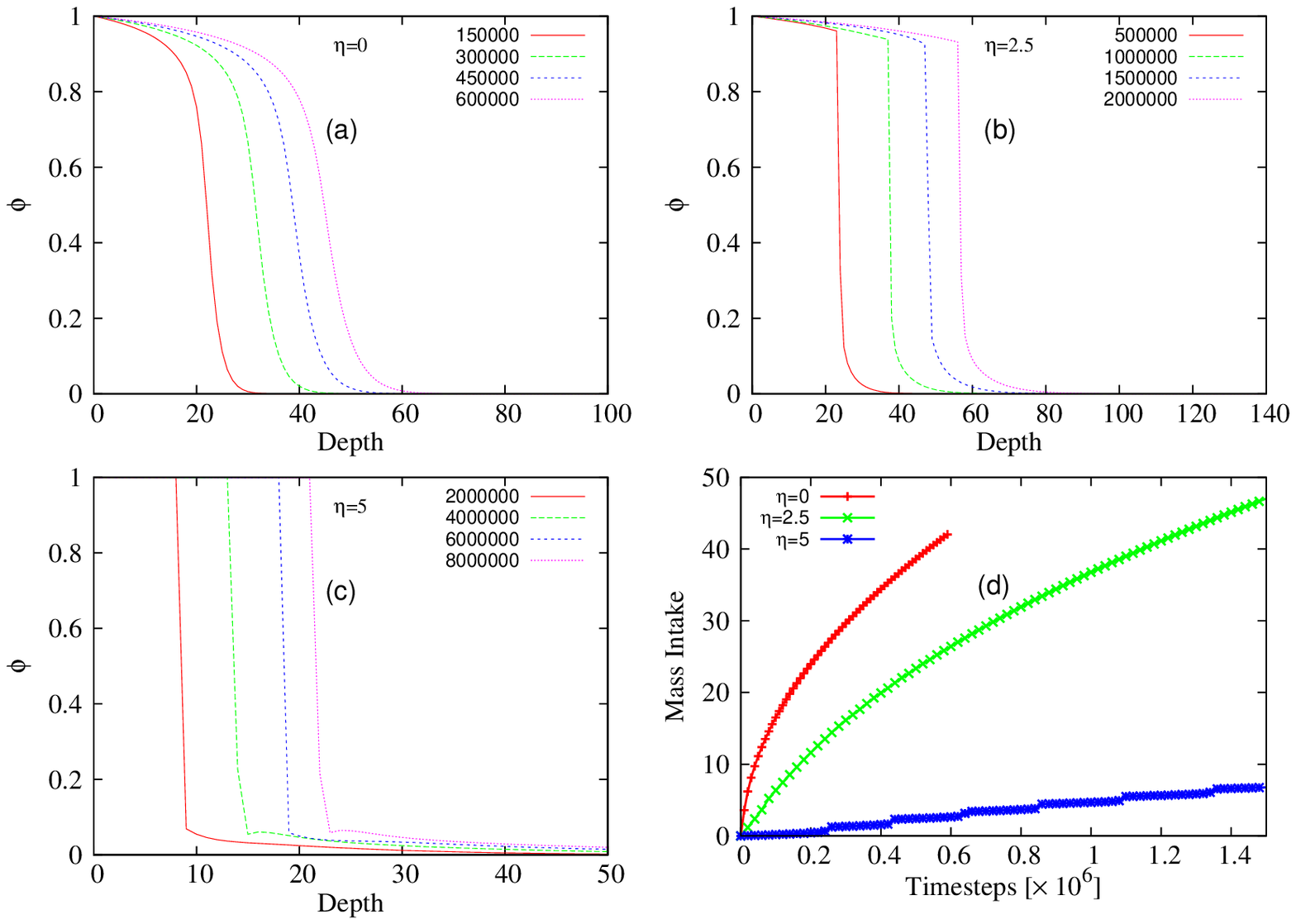}
\caption{Calculations performed with $\gamma$=5.0. (a)-(c): Concentration profiles at equally spaced time steps when $\eta=0, 2.5, 5.0$, respectively. (d): Mass intake as a function of time for different values of $\eta$: $0, 2.5, 5.0$.}
\label{numanomalous0}
\end{center}
\end{figure*}

\begin{figure*}[!htb]
\begin{center}
\includegraphics[scale=0.85]{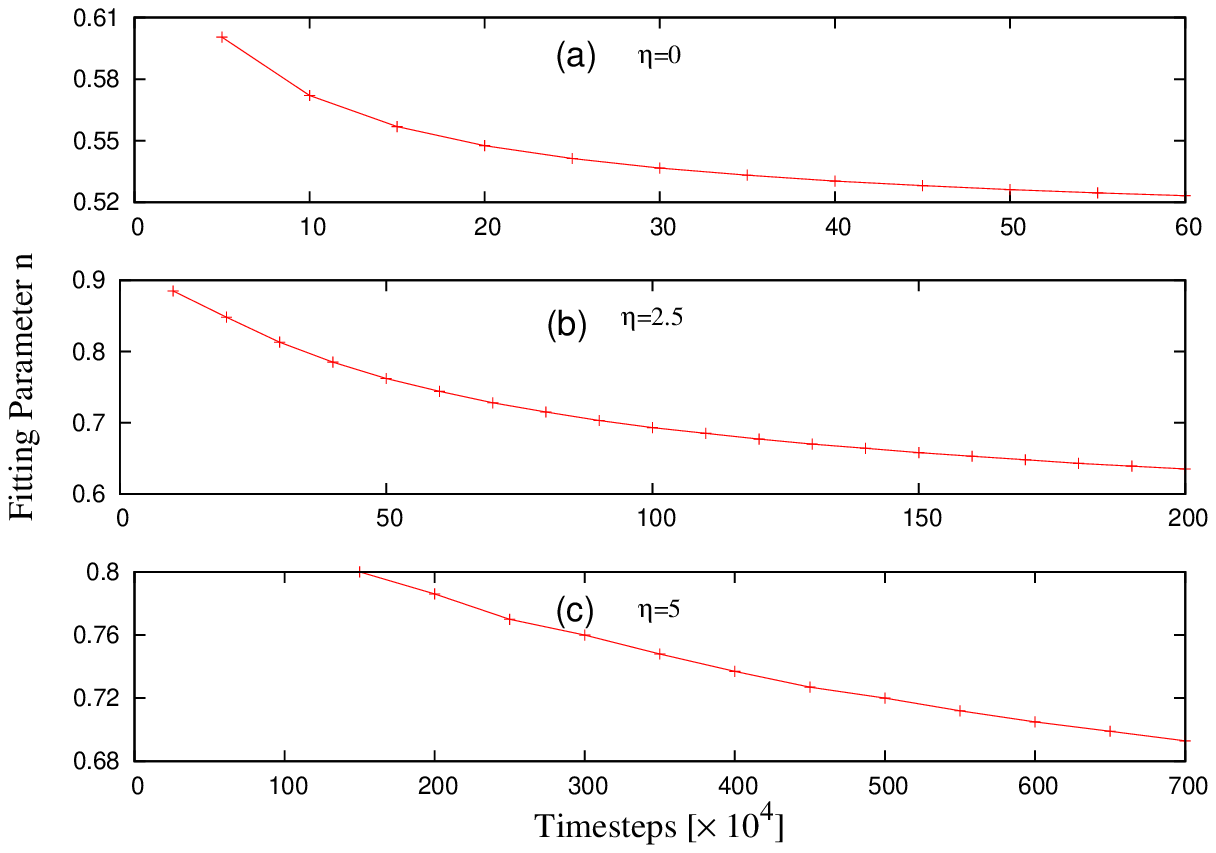}
\caption{The variation of fitting parameter $n$ with respect to diffusion time $t$ for different values of $\eta$: $0, 2.5, 5.0$ when $\gamma = 5.0$.}
\label{anomalousntime}
\end{center}
\end{figure*}

When $\gamma=5.0$, the results are as shown in Fig.~\ref{numanomalous0}. We again fit the mass-time curves with the function $f(t)=kt^n$, giving the results shown in Fig.~\ref{anomalousntime}. We see that, although $n$ is decreasing with the diffusion time, it remains greater than 0.5, particularly for larger values of $\eta$. According to the classification, this is the regime of anomalous diffusion.

\begin{figure*}[!htb]
\begin{center}
\includegraphics[scale=0.85]{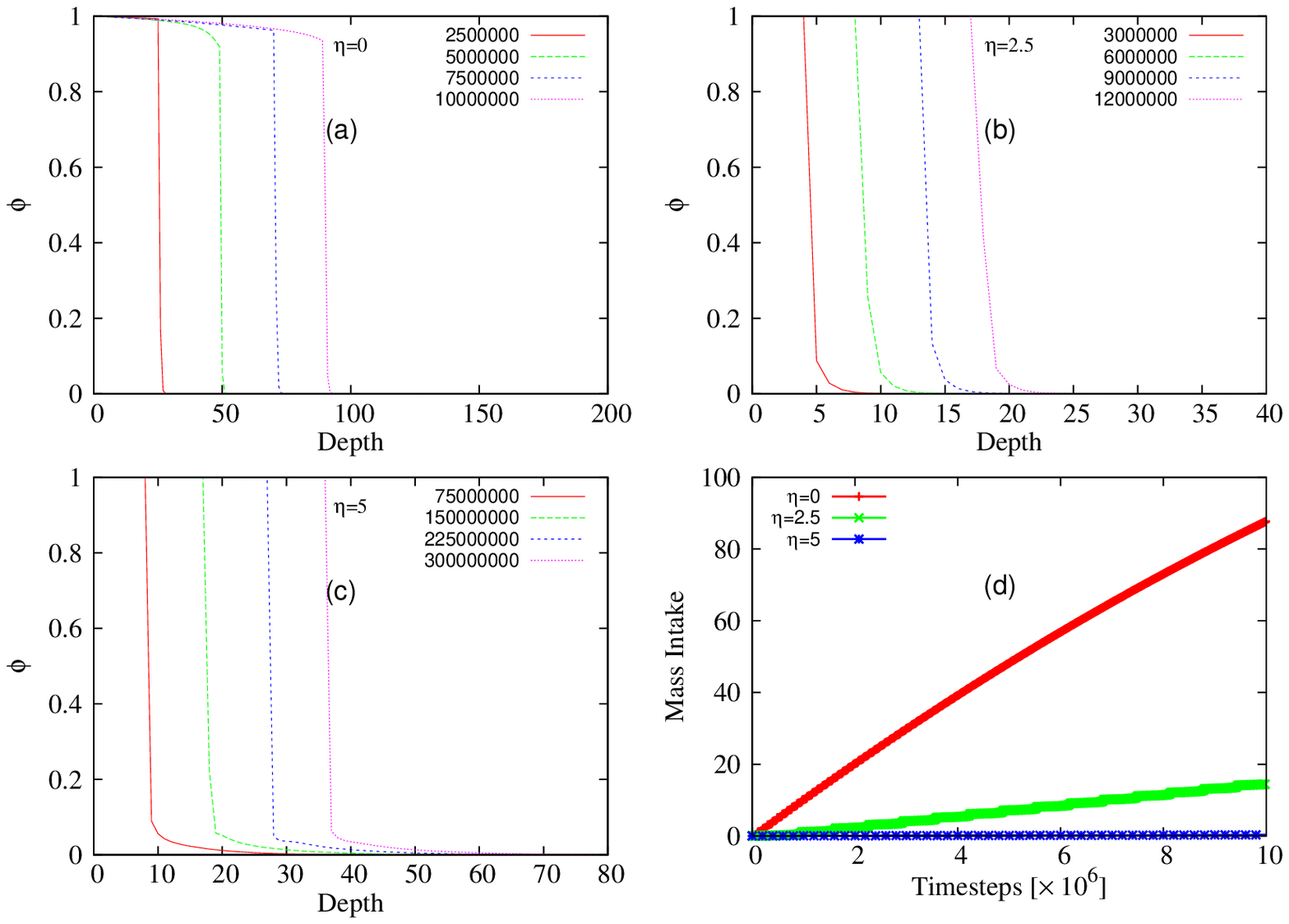}
\caption{Calculations performed with $\gamma$=9.6. (a)-(c): Concentration profiles at equally spaced time steps when $\eta=0, 2.5, 5.0$, respectively. (d): Mass intake as a function of time for different values of $\eta$: $0, 2.5, 5.0$.}
\label{numcase20}
\end{center}
\end{figure*}

\begin{figure}[!htb]
\begin{center}
\includegraphics[scale=0.85]{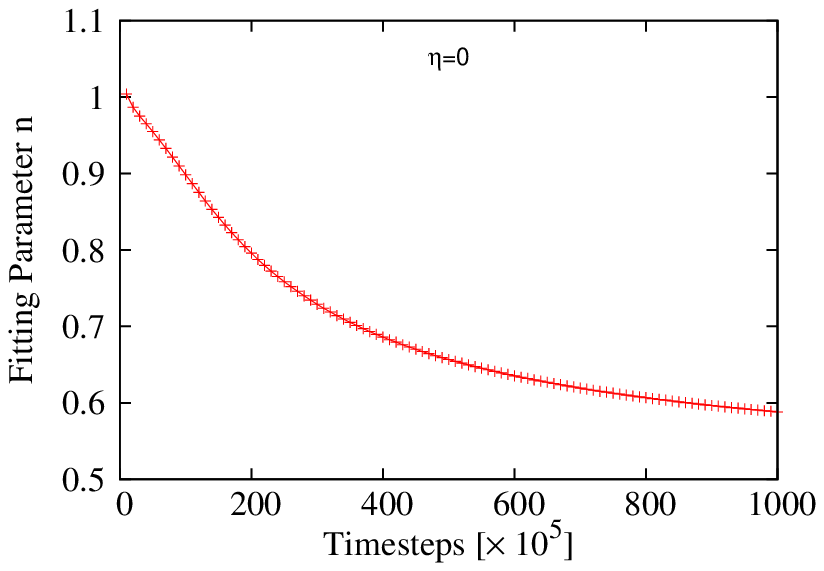}
\caption{The variation of fitting parameter $n$ with respect to diffusion time $t$ for $\gamma = 9.6$ and $\eta = 0$.}
\label{case2ntime}
\end{center}
\end{figure}

Finally, the results for $\gamma = 9.6$ are as shown in Fig.~\ref{numcase20}. We see that a sharp diffusion front develops in this case, which is a characteristic of Case II diffusion. Moreover, as $\eta$ becomes larger, a small increase in $\phi$ is seen to occur ahead of the sharp front. This phenomenon has been described as a ``Fickian precursor". \cite{qian} The mass intake is proportional to the time for all $\eta$ at small times. For $\eta = 0$, some curvature is visible at long times, but this is not observable in our results for $\eta = 2.5$ or $\eta = 5.0$, partly as a result of the very slow diffusion in the latter case. The increasing curvature of the mass-time curve as $t$ increases can be interpreted as a decrease in the parameter $n$ for the case where $\eta = 0$. This is illustrated in Fig.~\ref{case2ntime}, where $n$ is seen to drop from a value close to unity all the way down to 0.6 after 100 million time steps. Because we are assuming a sample with infinite thickness and all the reported experiments used very thin samples, this leads us to conclude that Case II diffusion is only a ``transient" phenomenon that happens at an early stage of diffusion under the above conditions. This has also been pointed out by Windle in a discussion of the limits of Case II sorption. \cite{windle} According to Fig.~\ref{case2ntime}, Case II diffusion happens in the first 3 million time steps for $\eta = 0$. The duration can be extended by applying a larger $\eta$. For example, when $\eta = 2.5$, the index $n$ remains as high as 0.992 after 15 million time steps; while for $\eta = 5$, the index $n$ remains as high as 0.996 after 300 million time steps. The duration can also be extended by increasing the exponent in the $\phi^4$ term in Eq.~(\ref{v2eq6}) to a higher value.

Super Case II diffusion happens when the two opposite sides of a sample are in contact with solvent. To examine this phenomenon we replace the boundary condition Eq.~(\ref{eq17}) by the requirement:
\begin{equation} \label{eq18}
\phi(x,t)|_{x=L}=1,
\end{equation}
where $L$ is the thickness of the sample. In our model, super Case II diffusion originates from the small energy difference $W$ in barrier height expressed by Eq.~(\ref{eq5}). When the two fronts originating from opposite sides of a simple slab approach each other, the swelling due to solvent molecules in one front propagates elastically into the vicinity of the other front. The parameter $\eta$ must necessarily be positive in order for this phenomenon to occur. 

Because the presence of a solvent molecule at site ${\boldsymbol l}$ will cause swelling that will be significant at sites more distant from ${\boldsymbol l}$ than the nearest-neighbor sites, it is necessary for the $W$ in our model to reflect this non-local influence. The barrier-height difference $W$ was represented in Eq.~(\ref{eq5}) as proportional to ${\boldsymbol a}\cdot\nabla\phi$, which was approximated in the previous calculation as $\phi_{l+a}-\phi_l$. For super Case II diffusion, this simple difference was replaced by a weighted average over a thickness of 1, 5, 9, and 13 layers, respectively. The weighting followed a binomial distribution, for which
\begin{equation} \label{eq19}
\delta\phi^{(1)}=\phi_{l+a}-\phi_{l}
\end{equation}
\begin{equation} \label{eq21}
\begin{split}
\delta\phi^{(5)}= &\frac{1}{16}[(\phi_{l+3a}-\phi_{l+2a})+4(\phi_{l+2a}-\phi_{l+a})+6(\phi_{l+a} \\
& -\phi_{l})+4(\phi_{l}-\phi_{l-a})+ (\phi_{l-a}-\phi_{l-2a})] \\
    = & \frac{1}{16}[\phi_{l+3a}+3\phi_{l+2a}+2\phi_{l+a}-2\phi_{l}-3\phi_{l-a}-\phi_{l-2a}]
\end{split}
\end{equation}
\begin{equation} \label{eq23}
\begin{split}
\delta\phi^{(9)}=& \frac{1}{256}[\phi_{l+5a}+7\phi_{l+4a}+20\phi_{l+3a}+28\phi_{l+2a}+14\phi_{l+a} \\
& -14\phi_{l}-28\phi_{l-a}-20\phi_{l-2a}-7\phi_{l-3a}-\phi_{l-4a}]
\end{split}
\end{equation}
\begin{equation} \label{eq24}
\begin{split}
\delta\phi^{(13)}=&\frac{1}{4096}[\phi_{l+7a}+11\phi_{l+6a}+54\phi_{l+5a}+154\phi_{l+4a} \\
& +275\phi_{l+3a}+297\phi_{l+2a}+132\phi_{l+a}-132\phi_{l}\\
& -297\phi_{l-a}-275\phi_{l-2a}-154\phi_{l-3a}-54\phi_{l-4a} \\
& -11\phi_{l-5a}-\phi_{l-6a}].
\end{split}
\end{equation}
The corresponding $\phi_{l+\frac{a}{2}}$ is also calculated using the same binomial distribution:
\begin{equation}
\phi_{l+\frac{a}{2}}^{(1)}=\frac{1}{2}(\phi_l+\phi_{l+a})
\end{equation}
\begin{equation}
\begin{split}
\phi_{l+\frac{a}{2}}^{(5)}=&\frac{1}{2}\times\frac{1}{16}[(\phi_{l+3a}+4\phi_{l+2a}+6\phi_{l+a}+4\phi_{l}+\phi_{l-a}) \\
& +(\phi_{l+2a}+4\phi_{l+a}+6\phi_{l}+4\phi_{l-a}+\phi_{l-2a})] \\
=&\frac{1}{32}(\phi_{l+3a}+5\phi_{l+2a}+10\phi_{l+a}+10\phi_{l}+5\phi_{l-a} \\
& +\phi_{l-2a})
\end{split}
\end{equation}
\begin{equation}
\begin{split}
\phi_{l+\frac{a}{2}}^{(9)}=&\frac{1}{512}(\phi_{l+5a}+9\phi_{l+4a}+36\phi_{l+3a}+84\phi_{l+2a}+126\phi_{l+a} \\
& +126\phi_{l}+84\phi_{l-a}+36\phi_{l-2a}+9\phi_{l-3a}+\phi_{l-4a})
\end{split}
\end{equation}
\begin{equation}
\begin{split}
\phi_{l+\frac{a}{2}}^{(13)}=&\frac{1}{8192}(\phi_{l+7a}+13\phi_{l+6a}+78\phi_{l+5a}+286\phi_{l+4a} \\
& +715\phi_{l+3a}+1287\phi_{l+2a}+1716\phi_{l+a}+1716\phi_{l} \\
&+1287\phi_{l-a}+715\phi_{l-2a}+286\phi_{l-3a}+78\phi_{l-4a} \\
&+13\phi_{l-5a}+\phi_{l-6a}).
\end{split}
\end{equation}

\begin{figure*}[!htb]
\begin{center}
\includegraphics[scale=0.85]{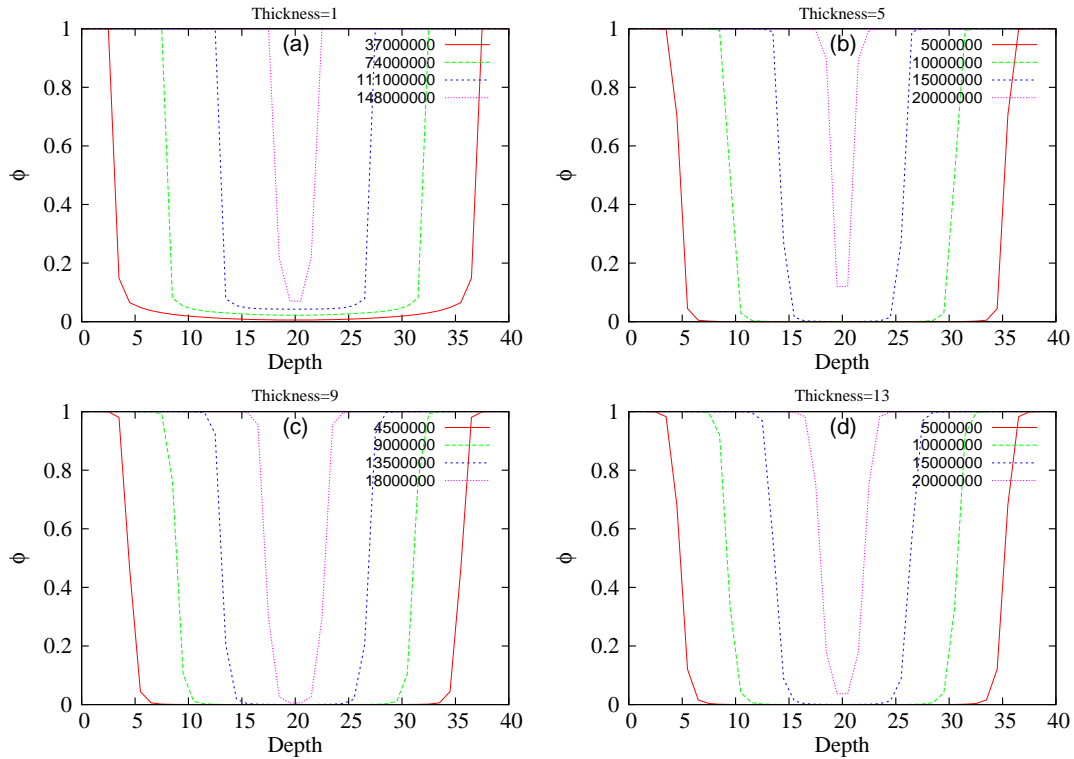}
\caption{Calculations performed with $\gamma$=9.6, and $\eta=5.0$ for a sample slab with its two opposite surfaces in contact with solvent. (a)-(d): Concentration profiles at equally spaced time steps when thickness = 1, 5, 9, 13, respectively.}
\label{numsupercase20}
\end{center}
\end{figure*}

\begin{figure}[!htb]
\begin{center}
\includegraphics[scale=0.85]{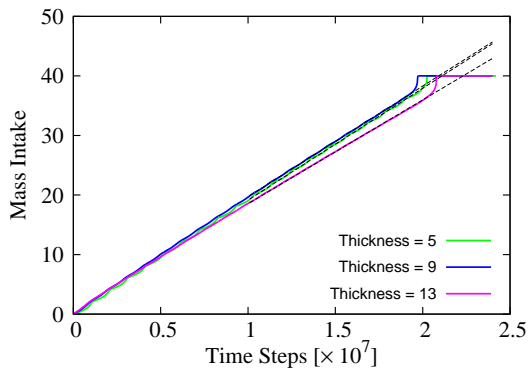}
\caption{Mass intake as a function of time for a sample slab with its two opposite surfaces in contact with solvent. Average thicknesses = 5, 9, 13, respectively. The trends of linear sorption are shown as dashed lines. These results resemble the Figures 2 and 3 in Jacques {\em et. al.}'s experimental work. \cite{jac}}
\label{numsupercase21}
\end{center}
\end{figure}

The resulting concentration profiles are shown in Fig.~\ref{numsupercase20}. We see that with the increase of chosen thickness, the ``precursor" becomes smaller, and the diffusion speed becomes larger. 
The corresponding curves of mass intake versus time are shown in Fig.~\ref{numsupercase21} for layers of thickness 5, 9, and 13 (the single-layer case is omitted, as no super Case II behavior is observable). In this figure, the trends of linear sorption are shown as dashed lines. For the three larger thicknesses considered, the acceleration in the rate of mass intake remains constantly observable in this model. This suggests that the particular phenomenon of super Case II diffusion requires the inclusion of longer-range elasticity effects for an accurate description. 

Our overall conclusion in this section is that it is possible to reproduce all the observed types of diffusion by extending the well known diffusion equation. The time variation of the mass intake, however, deviates from a simple power law when the type of diffusion is anomalous or Case II.
 
\section{A stochastic model}

\subsection{Simulation details}

In the preceding section, diffusion was modeled as an instantaneous process in which the local transport of solvent was determined solely by the concentration gradient at that instant of time. It is, however, also possible that effects that are nonlocal in time could occur. The barrier to hopping between two sites could reflect a previous history of occupation if local solvent-induced swelling persists for a brief time even in the absence of solvent molecules.

\begin{figure}[!htb]
\begin{center}
\includegraphics[width=0.45\textwidth]{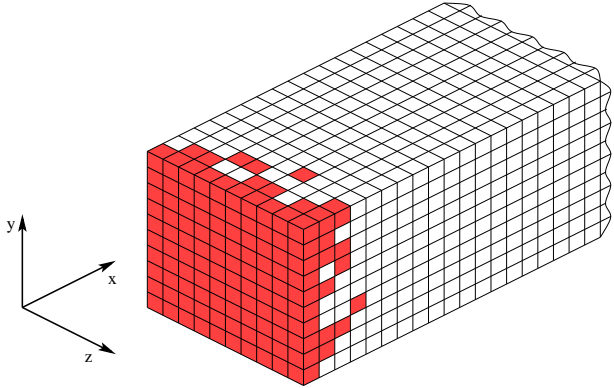}
\caption{A polymer bulk with cavities is considered to be a 3-dimensional grid. One surface of this grid is in contact with solvent.}
\label{3dgrid}
\end{center}
\end{figure}

To study whether such persistence effects would change the predictions made in the preceding sections, simulations were performed on a model in which the transition probabilities included a memory factor. The polymer network is represented by a three-dimensional grid as shown in Fig.~\ref{3dgrid}. The cubes in this grid denote cavities inside the polymer bulk, the walls between the cubes being polymer matrix. Transport is in the $x$-direction, and the number of cavities in each layer in the $y$-$z$ plane is set to 10$\times$10. Periodic boundary conditions are used in the $y$ and $z$ directions. During the simulation, the $x$ = 0 surface of this grid is considered to be in contact with the solvent, and accordingly all cavities on this surface are filled at all times. All other sites may be either empty or singly occupied. 

The diffusion occurs through a series of events in which a molecule in a cavity can transition only to one of its six nearest neighbors. To simulate the effect of previous history on the local transport, we characterize each wall as being either intact or ruptured, according to whether there has been a previous transit through that wall.

The probability of a transition occurring in which a solvent molecule moves between two sites will depend, as in the previous section, on the height of the barrier; this height, however, will now have a value that is dependent on past history rather than on the local concentration of solvent. We retain the language used in the previous section by denoting the barrier height of an intact wall as $V_b$, and that of a penetrated wall as $V_a$. The ratio of probabilities for transitions will then be $e^{-\beta V_b}/e^{-\beta V_a} = e^{-\gamma}$, where the dimensionless quantity $\gamma$ carries the same connotation as before, i.e., a measure of the barrier height reduction due to the presence of solvent. Now, however, the solvent may have been present in the past, but be currently absent. 

We must also take into account the tendency for the potential energy at a site to be lowered by the presence of solvent. This effect was characterized in the preceding section by the parameter $\eta$, which was proportional to $-dV_1/d\phi$. To introduce this effect into the simulation, we define $\phi_l\equiv\phi(x=l)$ as the average occupancy of the 100 sites within a slice of the grid in the $y$-$z$ plane. The probabilities of hopping in the positive-$x$ and negative-$x$ directions are then multiplied by a factor $e^{\eta\delta\phi_+}$ and $e^{-\eta\delta\phi_-}$, respectively, where $\delta\phi_+=\phi_{l+a}-\phi_l$ and $\delta\phi_-=\phi_l-\phi_{l-a}$.

The combined prescription for the transition probabilities is then as follows: In the forward direction the probability to make a transition is proportional to $e^{\eta\delta\phi}$ if the wall is unpenetrated, and $e^{\gamma+\eta\delta\phi}$ if it has been previously traversed; in the backward direction the probabilities are proportional to $e^{-\eta\delta\phi}$ for an unpenetrated wall and $e^{\gamma-\eta\delta\phi}$ for a penetrated one. Transitions at constant $x$ have no dependence on $\eta$, but the probability of occurrence contains the factor of $e^{\gamma}$ for travel through a ruptured wall. For each set of $\gamma$ and $\eta$, we perform simulations on 10 parallel systems, and then calculate ensemble averages to reduce the statistical error. 

 \subsection{Results for the stochastic model}
 
As in the preceding section, we find that the type of diffusion is mainly determined by $\gamma$ and mildly influenced by $\eta$. When $\gamma=0$, the diffusion is mainly Fickian; when $\gamma\gg1$, for example 9.6 in the following calculation, it is mainly Case II diffusion; and when $\gamma$ takes an intermediate value, such as 5.0, it is mainly anomalous diffusion. The major role of the parameter $\eta$ is its influence in producing a sharp diffusion front and in the formation of a precursor. 
 
 \subsubsection{Fickian diffusion}
 
When $\gamma=0$, the transition probability is not history dependent, and the probability of passing through a previously penetrated wall is the same as for an unpenetrated wall. 

\begin{figure*}[!htb]
\begin{center}
\includegraphics[scale=0.85]{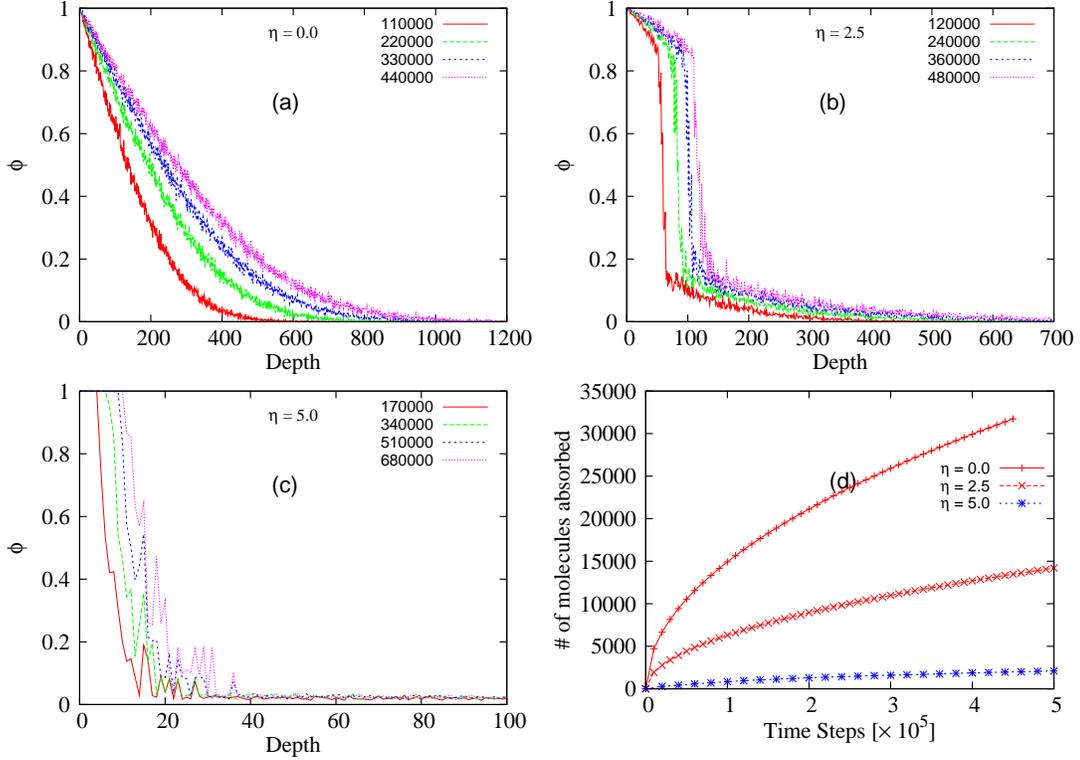}
\caption{(a)-(c): Simulated concentration profiles at equally spaced time steps for $\gamma = 0$ are shown at different values of $\eta$: 0, 2.5, 5.0. (d): Mass intake with respect to time at different values of $\eta$ when $\gamma = 0$.}
\label{fickian0}
\end{center}
\end{figure*}

The calculated concentration profiles at different times and different values of $\eta$ (0, 2.5, 5.0) are shown in Figs.~\ref{fickian0}(a)-(c). In these plots we see that, with increasing $\eta$, a sharp diffusion front begins to develop, as seen earlier in Fig.~\ref{numfickian0}. The plots when $\eta = 2.5$ in both Fig.~\ref{numfickian0} and Fig.~\ref{fickian0} resemble Fig. 8 of Thomas and Windle's paper, \cite{thom} where, in addition to the sharp front, there is also a pronounced precursor ahead of it, and a steep concentration gradient following it. 

Figure~\ref{fickian0}(d) shows the mass intake of solvent versus time at different values of $\eta$. These mass-time curves are then fitted with the function $f(t)=kt^n$, with $k$ and $n$ the fitting parameters. We find that, unlike the time local theory in Section II, the variation of $n$ with respect to time $t$ is greatly reduced after taking the non-local effect in time into account. The results for the exponent $n$ are shown in the second row of Table~\ref{table}. The mass-time curves generally follow a $t^{1/2}$ relation, which is a fingerprint of Fickian diffusion. When $\eta$ is larger than 5.0, the uncertainty in $n$ increases rapidly as a consequence of the slow diffusion speed and relatively short simulation time.

\begin{table}
\begin{center}
\begin{tabular}{  c | c | c | c }
    \hline
    $\eta$       &   0                  &  2.5             &  5.0                             \\ \hline
    $n$ for $\gamma = 0$             & 0.502     &  0.507  &  0.531         \\  \hline 
    $n$ for $\gamma = 5.0$          & 0.599            &  0.610       &    0.610      \\  \hline 
    $n$ for $\gamma = 9.6$          & 1.024              &    0.974          &   1.000             \\  \hline 
\end{tabular}
\end{center}
\caption{The fitting parameters found by fitting the function $f(t)=kt^n$ to the mass-time curves in Figs.~\ref{fickian0}(d) (second row), \ref{anomalous}(d) (third row), and \ref{case2}(d) (fourth row).}
\label{table}
\end{table}

 \subsubsection{Anomalous diffusion}

\begin{figure*}[!htb]
\begin{center}
\includegraphics[scale=0.85]{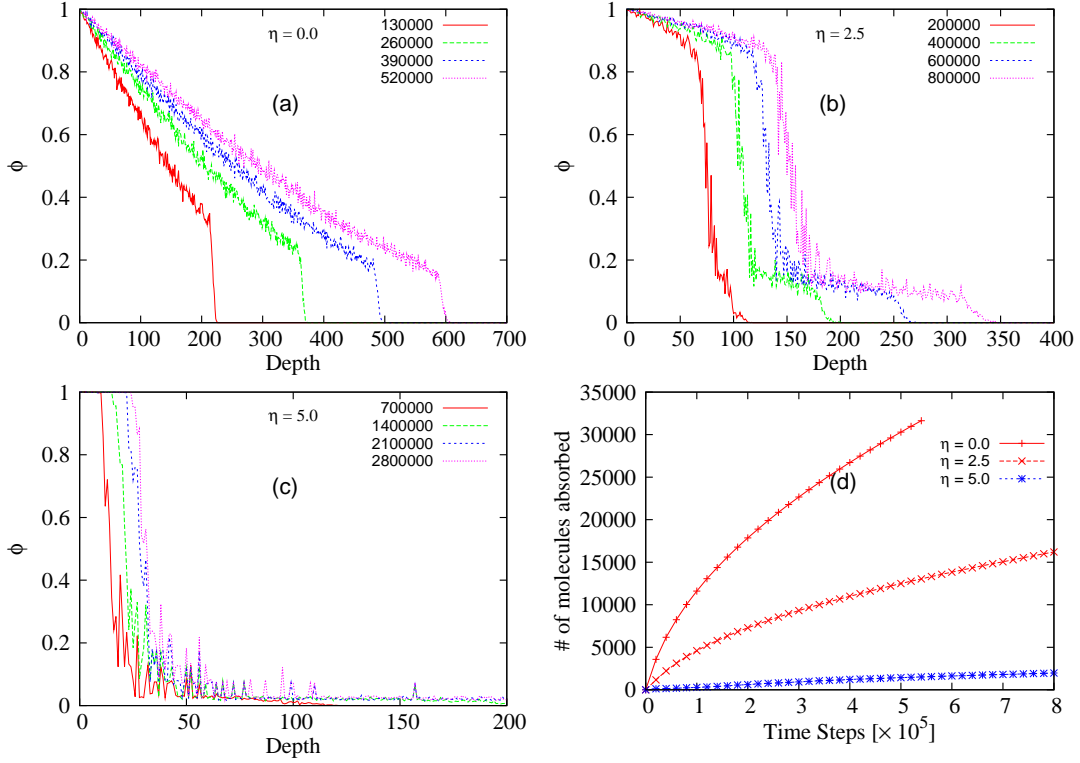}
\caption{(a)-(c): Simulated concentration profiles at equally spaced time steps for $\gamma = 5.0$ are shown at different values of $\eta$: 0, 2.5, 5.0. (d): Mass intake with respect to time at different values of $\eta$ when $\gamma = 5.0$.}
\label{anomalous}
\end{center}
\end{figure*}

When $\gamma=5.0$, the computed concentration profiles at different times and for different values of $\eta$ are shown in Figs.~\ref{anomalous}(a)-(c). In contrast with the Fickian diffusion discussed above, we notice the appearance of a sharp front at $\eta=0$. It is  also noticed that, with the increase of $\eta$, the concentration gradient behind the front becomes less steep, while the precursor ahead of the front penetrates  more deeply into the polymer matrix, although with reduced concentration range.

The mass-time curves are shown in Fig.~\ref{anomalous}(d); their fits with the function $f(t)=kt^n$ are shown in the third row of Table \ref{table}. We see that the exponents remain well above 0.5 with varying $\eta$. According to the present definition, this is the so-called ``anomalous" diffusion. 

 \subsubsection{Case II diffusion}

\begin{figure*}[!htb]
\begin{center}
\includegraphics[scale=0.85]{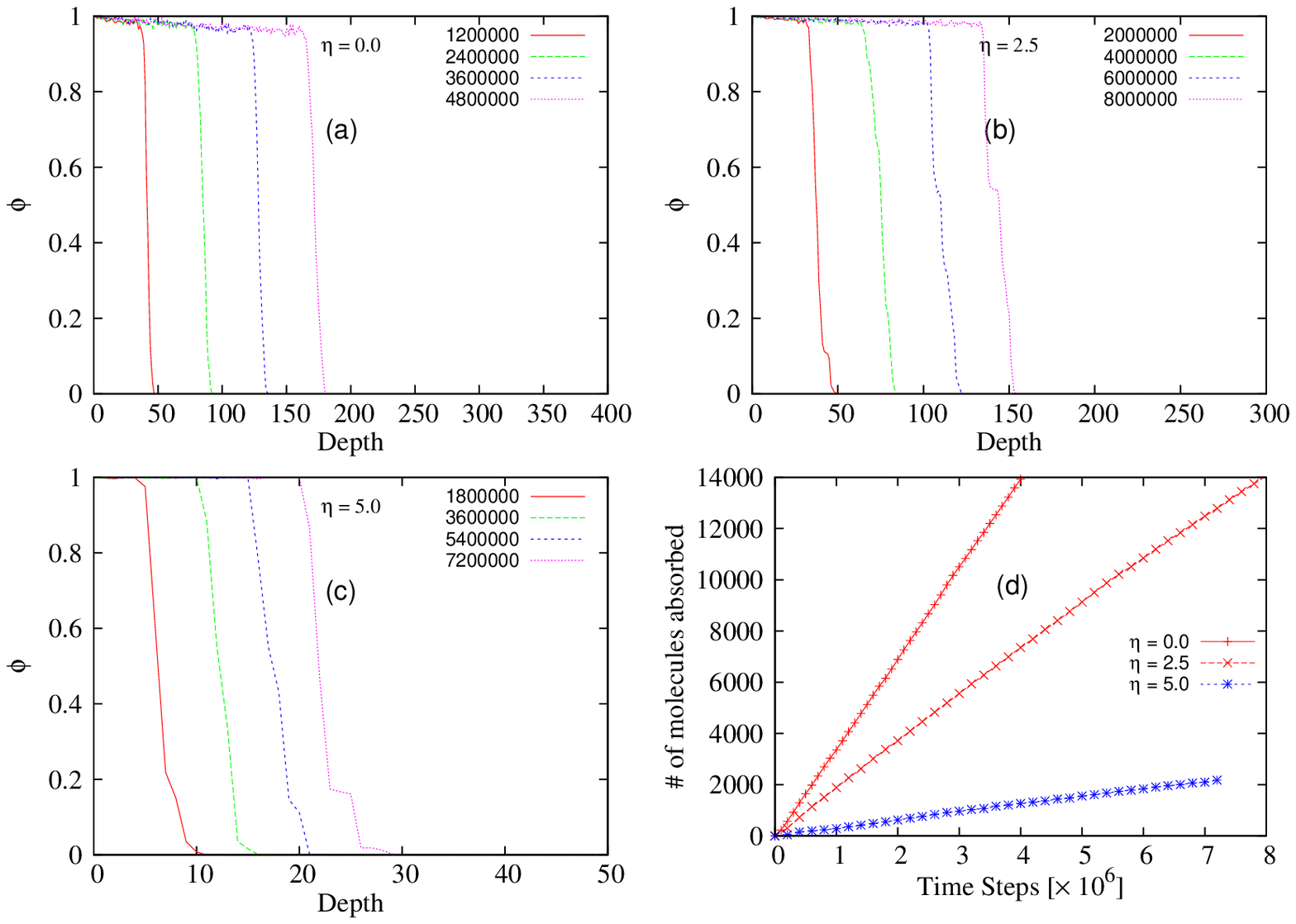}
\caption{(a)-(c): Simulated concentration profiles at equally spaced time steps for $\gamma = 9.6$ are shown at different values of $\eta$: 0, 2.5, 5.0. (d): Mass intake with respect to time at different values of $\eta$ when $\gamma = 9.6$.}
\label{case2}
\end{center}
\end{figure*}

When we choose $\gamma = 9.6$, the calculated concentration profiles at different times and values of $\eta$ are as shown in Figs.~\ref{case2}(a)-(c). We see that the sharp front now becomes pronounced, and the concentration gradient behind this front becomes small, which are the characteristic features of Case II diffusion. As $\eta$ becomes larger, the concentration gradient behind the front becomes less steep. 

The positions of concentration front at equally spaced time intervals in Figs.~\ref{case2}(a)-(c) indicate a nearly constant diffusion speed. This can be seen more clearly in Fig.~\ref{case2}(d), where the mass-time curves show that the amount of mass intake is proportional to time $t$, and the diffusion speed of the front is constant. The fitting parameter $n$ to the function $f(t)=kt^n$ is listed in the fourth row of Table \ref{table}. 

 \subsubsection{Super Case II diffusion}

As discussed in the theoretical analysis, the rate of mass uptake can rise rapidly when the precursors of the two fronts of diffusing solvent meet at the middle of a sample.

A grid of 20 layers was used to do the simulation.  During the simulation, both the $x=0$ and $x=L=19a$ surfaces of the grid were filled by solvent molecules at all times. The value of $\eta$ used here is 5.0, and $\gamma=9.6$. The approximated concentration gradient at position $x=l$ was calculated in the same way as in the previous theoretical section, according to Eqs.~(\ref{eq21}), (\ref{eq23}), and (\ref{eq24}). 

\begin{figure}[!htb]
\begin{center}
\includegraphics[scale=0.85]{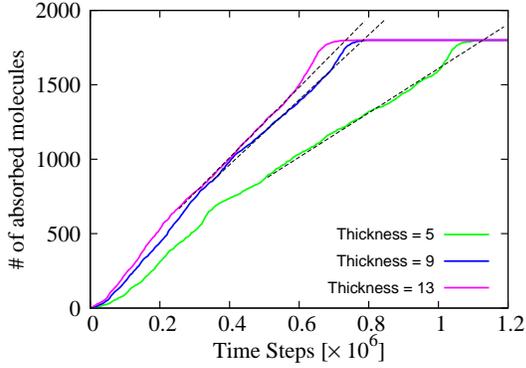}
\caption{Mass intake with respect to time for a sample slab with its two opposite surfaces in contact with solvent. The curve for ``Thickness = 5" is calculated according to Eq.~(\ref{eq21}), that for ``Thickness = 9" is calculated according to Eq.~(\ref{eq23}), and that for ``Thickness = 13" to Eq.~(\ref{eq24}). The value of $\eta$ used is 5.0. The trends of linear sorption are shown as dashed lines.}
\label{supercase2}
\end{center}
\end{figure}

The corresponding curves of mass intake versus time are shown in Fig.~\ref{supercase2} for layers of thickness of 5, 9, and 13. In the plot, where the trends of linear sorption are shown as dashed lines, we see an acceleration in the rate of mass intake at late times as the two fronts meet.

The conclusion from the simulations reported in this section is that our results from a model that is non-local in time are qualitatively similar to those of the time-local theory in Section II. One exception is that the slowing-down effect apparent in the theoretical results in Figs.~\ref{numanomalous0}(d) and \ref{numcase20}(d) is much less pronounced in the results of the simulations.

\section{The relation of our model to the Thomas-Windle model}

The widely discussed Thomas-Windle (TW) model can explain reasonably well the Case II diffusion. \cite{thom} To relate our model to the TW model, we develop our theoretical results a few steps further.
Using the relations
\begin{equation}
\begin{split}
\phi_{l+\frac{a}{2}}=\frac{1}{2}(\phi_{l+a}+\phi_l) \\
{\boldsymbol \nabla}\phi\cdot{\boldsymbol a}=\phi_{l+a}-\phi_l 
\end{split}
\end{equation}
Eq.~(\ref{v2eq7}) can be rewritten as:
\begin{equation}
\begin{split}
&\mbox{\boldmath $J$}_{{\boldsymbol l},{\boldsymbol l}+{\boldsymbol a}}=\mbox{\boldmath $a$} \nu e^{-\beta V_b} e^{\gamma \phi_{l+\frac{a}{2}}^4}\Bigl\{-{\boldsymbol \nabla}\phi\cdot{\boldsymbol a}\cosh(\eta{\boldsymbol \nabla}\phi\cdot{\boldsymbol a}) \\
&+\left[-2\phi_{l+\frac{a}{2}}^2+2\phi_{l+\frac{a}{2}}+\frac{1}{2}({\boldsymbol \nabla}\phi\cdot{\boldsymbol a})^2\right]\sinh(\eta{\boldsymbol \nabla}\phi\cdot{\boldsymbol a})\Bigr\}.
\end{split}
\end{equation}
We convert to a current density ${\boldsymbol j}$ by multiplying the density of sites, $\rho_0$, and transform ${\boldsymbol j}$ to take the form of Fick's first law, with 
\begin{equation} \label{diffusivity}
\begin{split}
D=&a^2\nu e^{-\beta V_b} e^{\gamma \phi^4}\Bigl\{\cosh(\eta{\boldsymbol \nabla}\phi\cdot{\boldsymbol a}) \\
&+\left[\frac{2\phi^2-2\phi}{{\boldsymbol \nabla}\phi\cdot{\boldsymbol a}}-\frac{1}{2}{\boldsymbol \nabla}\phi\cdot{\boldsymbol a}\right]\sinh(\eta{\boldsymbol \nabla}\phi\cdot{\boldsymbol a})\Bigr\},
\end{split}
\end{equation}
where we have dropped the subscript $l+\frac{a}{2}$ for $\phi$. We see that $D$ is a function of both $\phi$ and ${\boldsymbol \nabla}\phi$.
When $\eta$ is small ($\eta<0.5$), expansion of Eq.~(\ref{diffusivity}) to terms linear in ${\boldsymbol \nabla}\phi$ gives
\begin{equation}  \label{simd}
D=a^2\nu e^{-\beta V_b} e^{\gamma \phi^4}\left[1+2\eta\phi(\phi-1) \right],
\end{equation}
where all terms in ${\boldsymbol \nabla}\phi$ cancel out, and $D$ returns to being a function of $\phi$ only.

According to the TW model, \cite{thom} the rate of change of volume fraction of penetrant at each very thin material element is
\begin{equation} \label{eqb1}
\frac{\partial \phi}{\partial t}=\frac{P}{\eta_T k_c},
\end{equation}
where $k_c$ is a constant of proportionality. Here $P$ is osmotic pressure, which can be expressed as 
\begin{equation} \label{eqb2}
P=\frac{k_BT}{\Omega}ln\left(\frac{a_T}{\phi}\right),
\end{equation}
where $k_B$ is the Boltzmann constant, $\Omega$ is the partial molecular volume of penetrant, and $a_T$ is the penetrant activity. The quantity $\eta_T$ is the viscosity of the polymer, and is represented as 
\begin{equation} \label{eqb3}
\eta_T=\eta_{T0} e^{-M\phi},
\end{equation}
where M is a constant, and $\eta_{T0}$ is the viscosity of the unswollen polymer. (Note: the $\eta_T$ here has a different physical meaning from the $\eta$ in our theoretical model.)
The diffusion process is governed by the equation
\begin{equation} \label{eqb4}
\frac{\partial \phi}{\partial t}=\frac{\partial}{\partial x}\left(D^*(\phi)\frac{\phi}{a_T}\frac{\partial a_T}{\partial x}\right).
\end{equation}
By integrating the coupled Eqs.~(\ref{eqb1}) and (\ref{eqb4}) numerically with the extra assumption that $a_T$ is a state function of $\phi$, \cite{hui2} i.e., $a_T=a_T\left(\phi(x,t)\right)$, Thomas and Windle were able to reproduce the main features of Case II diffusion.
From Eq.~(\ref{eqb2}), we have 
\begin{equation}  \label{eqb5}
\frac{a_T}{\phi}=e^{\frac{P\Omega}{k_{B}T}},
\end{equation}
and thus
\begin{equation}  \label{eqb6}
\frac{d a_T}{d \phi}=e^{\frac{P\Omega}{k_{B}T}}\left(1+\frac{\phi\Omega}{k_BT}\frac{\partial P}{\partial \phi}\right).
\end{equation}
We also have
\begin{equation}  \label{eqb7}
\frac{\partial a_T}{\partial x}=\frac{d a_T}{d \phi}\frac{\partial \phi}{\partial x}.
\end{equation}
Substituting Eq.~(\ref{eqb6}) into (\ref{eqb7}), and then substituting Eqs.~(\ref{eqb5}) and (\ref{eqb7}) into Eq.~(\ref{eqb4}) yields
\begin{equation}  \label{eqb8}
\frac{\partial \phi}{\partial t}=\frac{\partial}{\partial x}\left[D^*(\phi)\left(1+\frac{\phi\Omega}{k_BT}\frac{\partial P}{\partial \phi}\right)\frac{\partial \phi}{\partial x}\right].
\end{equation}  
Because $a_T$ is a state function of $\phi$, according to Eq.~(\ref{eqb2}), $P$ is also a state function of $\phi$: $P=P(\phi)$. If we define a new entity, $D$, as 
\begin{equation} \label{eqb9}
D\equiv D^*(\phi)\left(1+\phi\Omega\beta\frac{\partial P}{\partial \phi}\right),
\end{equation}
where we have used $\beta=1/k_BT$, then $D$ is also a state function of $\phi$: $D=D(\phi)$. 
From the comparison with Eq.~(\ref{simd}), we have
\begin{equation} \label{eqb11}
a^2\nu e^{-\beta V_b} e^{\gamma \phi^4}\sim D^*(\phi)
\end{equation}
\begin{equation}
\eta\sim\frac{\Omega\beta}{2(\phi-1)}\frac{\partial P}{\partial \phi}.
\end{equation}
Thus the parameter $\eta$ is related to the osmotic pressure $P$ in the TW model. The quantities $\eta$ and $P$ play a similar role in each model. In our model, we simply assume $\eta$ to be constant, and mainly analyze the role played by $D^*(\phi)$, while the TW approach calculates $P$ by combining Eq.~(\ref{eqb1}) and Eq.~(\ref{eqb3}) to find
\begin{equation}
P=\eta_{T0}e^{-M\phi}k_c\frac{\partial \phi}{\partial t},
\end{equation}
and fixes the form of $D^*(\phi)$ to an exponential function of the concentration $\phi$. We note that if $P$ is a state function of $\phi$, then $\frac{\partial \phi}{\partial t}$ is also state function of $\phi$ only, which puts some constraint on the functional form of $\phi$.

\section{Relation to Experimental Data}

Now let us apply our model to some realistic systems. We choose the benzene-polystyrene system and the toluene-polystyrene system as examples. To calculate $\gamma$, we need to know the potential barrier in the absence of solvent ($V_b$) and at saturation ($V_a)$. These effective potential barrier heights are often referred to as the ``activation energy" $E_{\rm D}$ for diffusion. \cite{braune} Experimental results for the activation energy at these two extreme concentrations are sparse, and so we can only make an estimate based on limited sources. \cite{zielinski,vrentas} The detailed analysis leading to this estimation is in the Appendix. It is found that for the benzene-polystyrene system at 30 $^\circ$C, $V_b\simeq4.86\times10^{-20}$ J, and  $V_a\simeq2.78\times10^{-20}$ J. Thus, $\gamma=\beta(V_b-V_a)=(4.86-2.78)\times10^{-20}\rm{J}/(1.38\times10^{-23}\rm{J/K}\times303\rm{K})\simeq5.0$. According to our model, this indicates that the diffusion of benzene into polystyrene is anomalous diffusion, which is in accord with the results reported by Long and Kokes. \cite{long} For the toluene-polystyrene system at 30 $^\circ$C, we have $V_b\simeq9.24\times10^{-20}$ J, and  $V_a\simeq2.57\times10^{-20}$ J. Then we have $\gamma=\beta(V_b-V_a)=(9.24-2.57)\times10^{-20}\rm{J}/(1.38\times10^{-23}\rm{J/K}\times303\rm{K})\simeq15.9$, which indicates that the diffusion of toluene into polystyrene should be Case II type according to our model. Experiments by Gall and Kramer for the diffusion of deuterated toluene in polystyrene led to the same conclusion. \cite{gall}

We have seen in the preceding sections that the Taylor expansion of Eq.~(\ref{diffusivity}) at small $\eta$ leads to a concentration-dependent diffusivity, which can also lead to all types of diffusion. Now we give a theoretical derivation of how to verify the existence of a concentration-dependent diffusivity using the diffusion profile observed in experiments. 
From the non-linear diffusion equation
\begin{equation}
\frac{\partial \phi}{\partial t}=\frac{\partial}{\partial x}\left(D(\phi)\frac{\partial \phi}{\partial x} \right),
\end{equation}
we have 
\begin{equation}
\begin{split}
\int_x^\infty\frac{\partial}{\partial t} \phi(x^\prime,t)dx^\prime=\left[D\frac{\partial \phi}{\partial x}\right]_x^\infty=-D(\phi)\frac{\partial}{\partial x}\phi(x,t).
\end{split}
\end{equation}
So 
\begin{equation} \label{profiletod}
D(\phi)=-\frac{\int_x^\infty\frac{\partial}{\partial t} \phi(x^\prime,t)dx^\prime}{\frac{\partial}{\partial x}\phi(x,t)}\equiv g(x,t).
\end{equation}
Inverting $\phi=\phi(x , t)$ to $x=x(\phi , t)$, we have 
\begin{equation} \label{dphi}
D(\phi)=g[x(\phi, t), t].
\end{equation}

Ideally, from the experimentally observed concentration profiles, we could calculate $g[x(\phi, t), t]$ according to Eq.~(\ref{profiletod}). If $g[x(\phi, t), t]$ is independent of $t$, then the diffusivity $D(\phi)$ is concentration dependent only, and provides a complete description of diffusion. Unfortunately, the precision of most measurements is not sufficient to achieve this goal. The relevant data can be found in the work of Hermes {\em et. al.} \cite{hermes} and Ogieglo {\em et. al.} \cite{ogi1, ogi2}

\section{Summary and Conclusions}

In this paper we have set up a random walker model to theoretically analyze and a three-dimensional-grid model to simulate the diffusion behavior of small solvent molecules in a glassy polymer matrix. In the theoretical analysis, we generalized the diffusion equation, and then solved it numerically. Results showed that the diffusion type was mainly determined by a dimensionless parameter $\gamma \equiv \beta(V_b-V_a)$, where $V_b$ is the potential barrier between cavities in the polymer matrix in the absence of solvent ($\phi=0$), and $V_a$ is the lowered barrier at saturation ($\phi=1$). In one limit, when $\gamma = 0$, the diffusion is Fickian, while in another limit, when $\gamma \gg 1$, the diffusion becomes Case II type. When $\gamma$ takes intermediate values, it is anomalous diffusion. Super Case II diffusion, which happens when the two opposite sides of a sample are in contact with solvent, originates from the swelling due to solvent molecules in one front propagating elastically into the vicinity of the other front. Using the same assumptions as that in the theoretical model, with an extra memory factor reflecting a previous occupation history of cavities, we performed simulations, and obtained similar results.

We also derived the formal expression of diffusivity by comparing with Fick's first law. Generally speaking, the diffusivity depends on $\phi$ and ${\boldsymbol \nabla}\phi$; but when another dimensionless parameter $\eta$ (which is a measure of the variation of the base potential energy $V_1$ in the polymer matrix with respect to solvent concentration $\phi$) in our model is small, the dependence of $D$ on ${\boldsymbol \nabla}\phi$ vanishes, and the expression for $D$ becomes a function of $\phi$ only. At this limit the Thomas-Windle model can be related and incorporated into our model.

According to our model, the immense range of diffusion coefficients comes from polymer matrix properties (the average distance between cavities $a$, the height of potential barriers between cavities $V_a, V_b$), environment (temperature $T$), and solvent properties (attempt frequency $\nu$). However, only the relative magnitudes of the diffusion coefficients determine the diffusion type, while their absolute magnitudes only influence the time scale at which diffusion takes place. The present analysis provides an overview that incorporates the many specialized models interpreting specific types of diffusion \cite{qian,thom, tsige1, tsige2, ferreira} into one unified model.

\begin{acknowledgments}
This work was supported by the National Science Foundation under Grant DMR-1410290 (M.T.) and by the Petroleum Research Fund of the American Chemical Society under Grant PRF\# 51995-ND7 (J.M.\&P.L.T.), and was made possible by use of facilities at the Case ITS High Performance Computing Cluster.

\end{acknowledgments}

\appendix*

\section{Estimation of the activation energy}

The effective potential barrier height defined in Eq.~(\ref{eqv0}) is often referred to as the ``activation energy" $E_{\rm D}$ for diffusion. Our work shows that the activation energy for non-Fickian diffusion is concentration-dependent. Duda {\it et al.} \cite{zielinski} calculated the concentration dependence of the activation energy $E$ required to overcome attractive forces in the Vrentus-Duta free-volume diffusion model, according to which the self diffusion coefficient is given by: \cite{vrentas}
\begin{widetext}
\begin{equation} \label{d_free_v}
D_1=D_0\exp\left(\frac{-E}{RT}\right)\exp\left[\frac{-(\omega_1\hat{V}_1^*+\omega_2\xi \hat{V}_2^*)}{\omega_1\left(\frac{K_{11}}{\gamma^*}\right)(K_{21}-T_{g1}+T)+\omega_2\left(\frac{K_{12}}{\gamma^*}\right)(K_{22}-T_{g2}+T)}\right],
\end{equation}
\end{widetext}
where $D_0$ is a constant pre-exponential factor; $E$ is the energy per mole that a solvent molecule needs to overcome the attractive forces that constrain it to its neighbors; $R$ is the ideal gas constant; $\omega_i$ is the mass fraction for component $i$ ($i$ equals to 1 or 2, 1 for solvent and 2 for polymer); $\hat{V}_i^*$ is the specific critical hole free volume required for jump for component $i$; $T_{gi}$ is the glass transition temperature for component $i$; $\xi$ is the ratio of molar volumes for the solvent and polymer jumping units; $\gamma^*$ is an overlap factor (between $1/2$ and 1) introduced because the same free volume is available to more than one molecule, and thus has a different physical meaning from the $\gamma$ in our present model; $K_{11}$ and $K_{21}$ are free-volume parameters for the solvent; $K_{12}$ and $K_{22}$ are free-volume parameters for the polymer. The activation energy for diffusion $E_{\rm D}$, on the other hand, can be defined according to the Arrhenius equation: 
\begin{equation}
D_1=D_0\exp\left(\frac{-E_{\rm D}}{RT}\right),
\end{equation}
which finally gives us
\begin{equation} \label{activation_e}
E_{\rm D}=-R\frac{\partial \ln D_1}{\partial(1/T)}=RT^2\frac{\partial \ln D_1}{\partial T}.
\end{equation}
Substituting Eq.~(\ref{d_free_v}) into Eq.~(\ref{activation_e}), we have the relation between $E_{\rm D}$ and $E$:
\begin{widetext}
\begin{equation} \label{ed_e_relation}
E_{\rm D}=E+RT^2\left\{\frac{(\omega_1\hat{V}_1^*+\omega_2\xi\hat{V}_2^*)\left(\omega_1\frac{K_{11}}{\gamma^*}+\omega_2\frac{K_{12}}{\gamma^*}\right)}{\left[\omega_1\frac{K_{11}}{\gamma^*}(K_{21}-T_{g1}+T)+\omega_2\frac{K_{12}}{\gamma^*}(K_{22}-T_{g2}+T)\right]^2} \right\}.
\end{equation}
\end{widetext}

\begin{figure*}[!htbp]
\begin{center}
\includegraphics[scale=0.85]{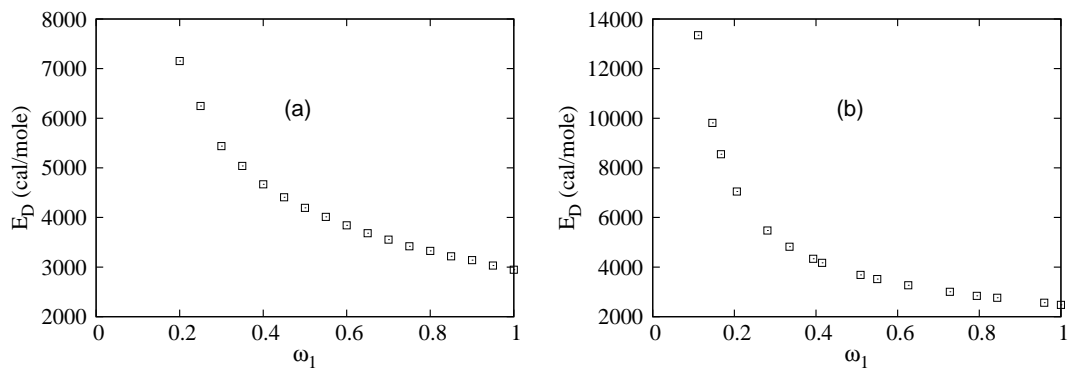}
\caption{The dependence of activation energy $E_{\rm D}$ on solvent concentration $\omega_1$ for the benzene-polystyrene system (a) and toluene-polystyrene system (b).}
\label{e_omega_relation}
\end{center}
\end{figure*}

Now let us calculate the $E_{\rm D}$ for self diffusion of benzene in polystyrene at 30 $^\circ$C. The values of parameters for this system are: \cite{zielinski} $\hat{V}_1^*=0.901$ cm$^3$/g, $\hat{V}_2^*=0.850$ cm$^3$/g, $K_{11}/\gamma^*=1.48\times10^{-2}$ cm$^3$/(g$\cdot$K), $K_{12}/\gamma^*=5.82\times10^{-4}$ cm$^3$/(g$\cdot$K), $K_{21}-T_{g1}=-178.52$ K,  $K_{22}-T_{g2}=-327.0$ K, $\xi=0.51$, $D_0=1.80\times10^{-3}$ cm$^2$/s. We also have $T=303.15$ K, $R=1.987$ cal/(mol$\cdot$K), and $\omega_2=1-\omega_1$. Using Eq.~(\ref{ed_e_relation}) and the $E$-$\omega_1$ relation in Fig.~4 of Duda {\it et al.}'s work, \cite{zielinski} we are able to obtain the $E_{\rm D}$-$\omega_1$ relation as shown in Fig.~\ref{e_omega_relation} (a).

One point to be noted is that the $\phi=1$ state in our model, when all the cavities are occupied, does not correspond to the $\omega_1=1$ state here, which stands for the pure benzene state, but should correspond to a smaller $\omega_1$, say, $\omega_1=0.5$. The activation energy at this point is about 4000 cal/mole. This gives an estimation of the value of $V_a$ in our model: $V_a\simeq4000\ \rm{cal/mole}=2.78\times10^{-20}$ J. The largest possible activation energy at low concentration can also be estimated from Fig.~\ref{e_omega_relation} (a), and is about 7000 cal/mole, which gives an estimation of the value of $V_b$: $V_b\simeq7000\ \rm{cal/mole}=4.86\times10^{-20}$ J.

Using the same method, we also calculated the $E_{\rm D}$-$\omega_1$ relation for the toluene-polystyrene system at 30 $^\circ$C as shown in Fig.~\ref{e_omega_relation} (b). The activation energy at $\phi=1$ can be estimated from the value of $E_{\rm D}$ at $\omega_1\simeq0.5$, which is about 3700 cal/mole. The estimation of the value of $V_a$ turns out to be: $V_a\simeq3700\ \rm{cal/mole}=2.57\times10^{-20}$ J. The largest possible activation energy at low concentration is estimated to be 13300 cal/mole. Then the corresponding value of $V_b$ is estimated to be: $V_b\simeq13300\ \rm{cal/mole}=9.24\times10^{-20}$ J.


\end{document}